\documentclass[%
 reprint,
 amsmath,amssymb,
 aps,
]{revtex4-2}

\usepackage{graphicx}% Include figure files
\usepackage{dcolumn}% Align table columns on decimal point
\usepackage{bm}% bold math
\usepackage[caption=false]{subfig}
\usepackage{siunitx, soul, xcolor}
\usepackage{adjustbox}

\let\oldst\st
\renewcommand{\st}[1]{{\textcolor{blue}{\oldst{#1}}}}
\begin{document}
\preprint{APS/123-QED}

\title{Epitaxial Growth and Domain Structure Imaging of\\ Kagome Magnet Fe$_3$Sn$_2$}

\author{Shuyu Cheng}
\thanks{These authors contributed equally.}
\affiliation{Department of Physics, The Ohio State University, Columbus, Ohio 43210, USA}
\author{Igor Lyalin}
\thanks{These authors contributed equally.}
\affiliation{Department of Physics, The Ohio State University, Columbus, Ohio 43210, USA}
\author{Alexander J. Bishop}
\affiliation{Department of Physics, The Ohio State University, Columbus, Ohio 43210, USA}
\author{Roland K. Kawakami}
\email{kawakami.15@osu.edu}
\affiliation{Department of Physics, The Ohio State University, Columbus, Ohio 43210, USA}

%\keywords{kagome ferromagnet, MBE, anomalous Nernst effect}

\begin{abstract}
Magnetic materials with kagome crystal structure exhibit rich physics such as frustrated magnetism, skyrmion formation, 
topological flat bands, and Dirac/Weyl points. Until recently, most studies on kagome magnets have been performed on bulk crystals or polycrystalline films. 
Here we report the synthesis of high-quality epitaxial films of topological kagome magnet Fe$_3$Sn$_2$ by atomic layer molecular beam epitaxy. 
Structural and magnetic characterization of Fe$_3$Sn$_2$ on epitaxial Pt(111) identifies highly ordered films with c-plane orientation and an in-plane magnetic easy axis.
Studies of the local magnetic structure by anomalous Nernst effect imaging reveals in-plane oriented micrometer size domains.
The realization of high-quality films by atomic layer molecular beam epitaxy opens the door to explore the rich physics of this system and investigate novel spintronic phenomena by interfacing Fe$_3$Sn$_2$ with other materials.

\end{abstract}

\flushbottom
\maketitle
\thispagestyle{empty}

 \begin{figure}
    \includegraphics[width=0.45\textwidth]{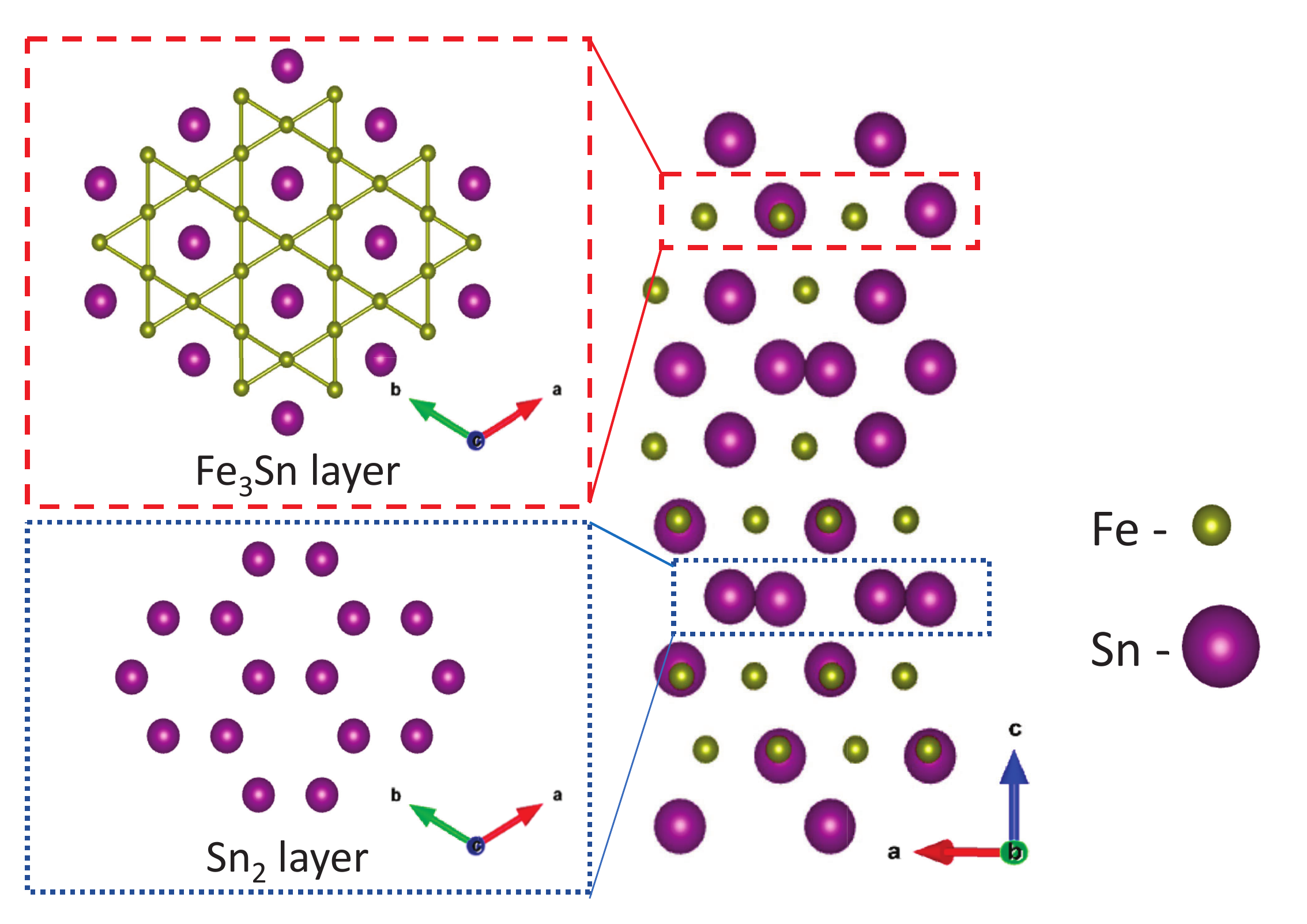}
    \caption{\label{fig:kagome} Left: Top view of an individual Fe$_3$Sn layer with kagome structure (top) and Sn$_2$ layer with honeycomb structure (bottom), respectively. Right: Side view of the crystal structure of Fe$_3$Sn$_2$ consisting of alternating stacking of two Fe$_3$Sn kagome layers and one Sn$_2$ layer.}
\end{figure}

In recent years, studies on magnetic topological materials with kagome lattices have become one of the hottest frontiers of condensed matter research, due to their exotic physical properties in both real space and momentum space~\cite{smejkal2018,yang2017topological}. 
In momentum space, angle-resolved photoemission spectroscopy (ARPES) experiments on Mn$_3$Sn, Fe$_3$Sn$_2$, FeSn, and CoSn~\cite{kuroda2017,ye2018,kang2020,kang2020topological} show that kagome lattices give rise to Dirac cones and flat bands which are topologically protected and are of particular interest.
In addition, scanning tunneling spectroscopy finds evidence for topological flat bands as a sharp peak in the local density of states~\cite{yin2019negative}.
These topologically nontrivial features result in signatures of anomalous transport (e.g. chiral anomaly) in magnetotransport experiments~\cite{kuroda2017,chen2021}.
Furthermore, it is theoretically predicted that the band structures of the kagome topological magnets can be controlled by tuning of their magnetic structures~\cite{kuroda2017,smejkal2018}.
In real space, the kagome topological magnets have layered structures with spins occupying corner-sharing triangular lattices, which leads to geometrical spin frustration~\cite{fenner2009non,nakatsuji2015}.
A surprisingly large anomalous Hall effect (AHE) and magneto-optic Kerr effect (MOKE) have been reported in noncollinear antiferromagnet Mn$_3$Sn, even with vanishingly small net magnetization~\cite{nakatsuji2015,higo2018}.
Skymrmion spin textures have been observed in ferromagnetic  Fe$_3$Sn$_2$ resulting from the competition of exchange, dipolar, and Zeeman energies~\cite{hou2017,hou2019}.
However, most of the studies on the kagome magnets have been done on bulk materials ~\cite{fenner2009,kida2011,nakatsuji2015,nayak2016,hou2017,ye2018,higo2018,hou2019,kang2020} with a few papers reporting the growth and characterization of epitaxial films~\cite{markou2018,inoue2019,taylor2020,khadka2020,hong2020molecular}.
Looking forward, epitaxial films will provide new opportunities for exploring fundamental physics and potential applications by tuning the dimensionality in thin films and interfacing Fe$_3$Sn$_2$ with different materials. 
In addition, preparing and visualizing a well-defined domain structure is crucial for studying the spatially resolved physics of kagome magnets, such as spin transfer torque and spin-orbit torque induced domain wall and skyrmion motion~\cite{parkin2008,fert2017}.
While the local magnetic structure of bulk Fe$_3$Sn$_2$ has been recently studied and room temperature magnetic skyrmions were revealed~\cite{hou2017,hou2019}, there have been no studies probing the domain structure of Fe$_3$Sn$_2$ in thin films.

In this paper, we report the epitaxial growth and magnetic domain imaging of high-quality Fe$_3$Sn$_2$ thin films on Pt(111)/Al$_2$O$_3$(0001) substrates. We utilize atomic layer molecular beam epitaxy (MBE) to synthesize Fe$_3$Sn$_2$ films by sequentially depositing Fe$_3$Sn kagome layers and Sn$_2$ layers (see Figure 1).
Structural characterization by \textit{in situ} reflection high energy electron diffraction (RHEED) and X-ray diffraction (XRD) 
confirm the crystalline structure of Fe$_3$Sn$_2$. 
The magnetic properties of Fe$_3$Sn$_2$ are investigated using MOKE and the anomalous Nernst effect (ANE) and consistently observe easy-plane magnetic anisotropy with square hysteresis loops.
Using a microscopy technique based on ANE, we successfully image the in-plane oriented domain structure of the epitaxial Fe$_3$Sn$_2$ films and investigate the magnetization reversal as a function of applied field.

\begin{figure}
    \subfloat[\label{fig:RHEED}]{
        \includegraphics[width=0.45\textwidth]{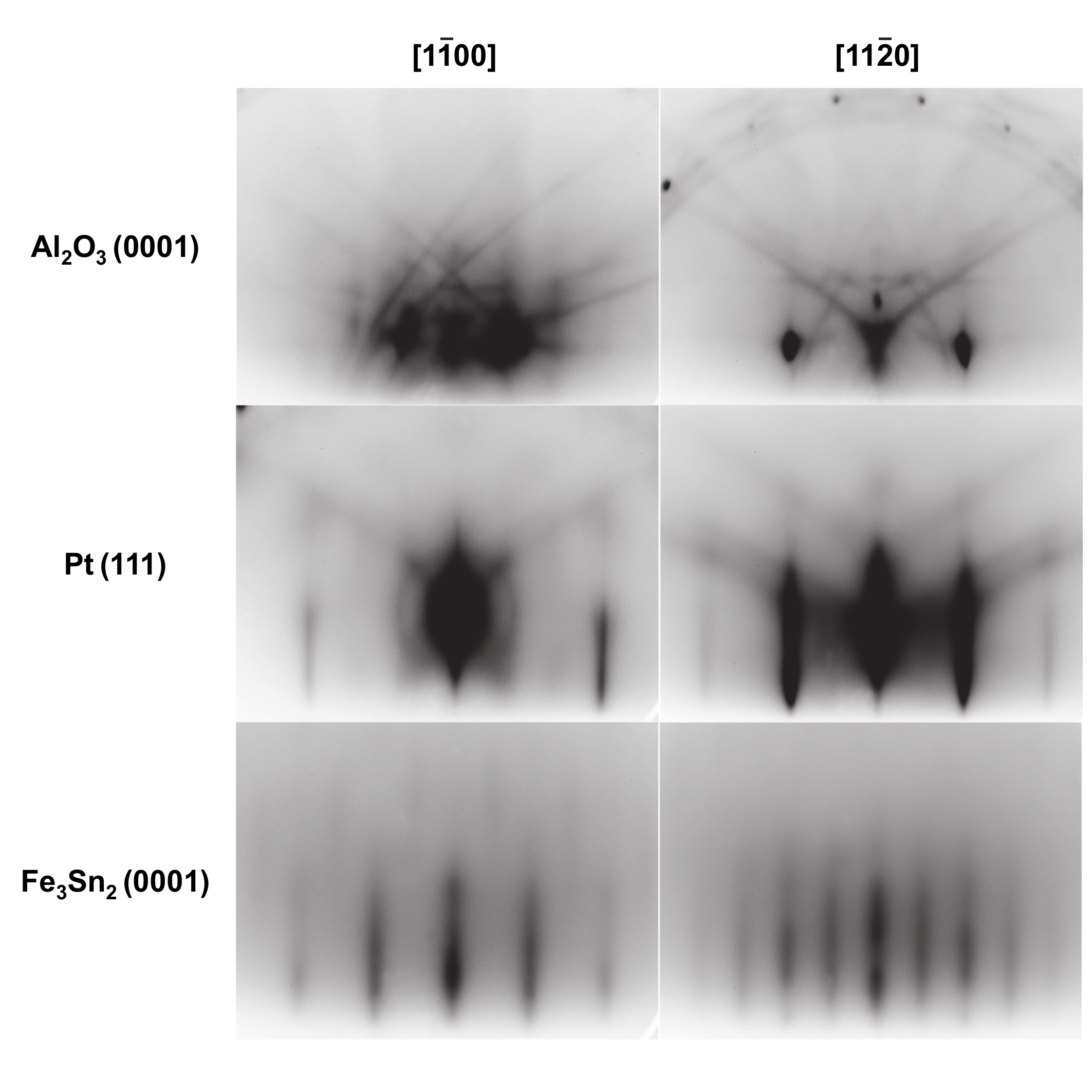}
    }

    \hfill
    \subfloat[\label{fig:RHEED_oscillations}]{
       \includegraphics[width=0.45\textwidth]{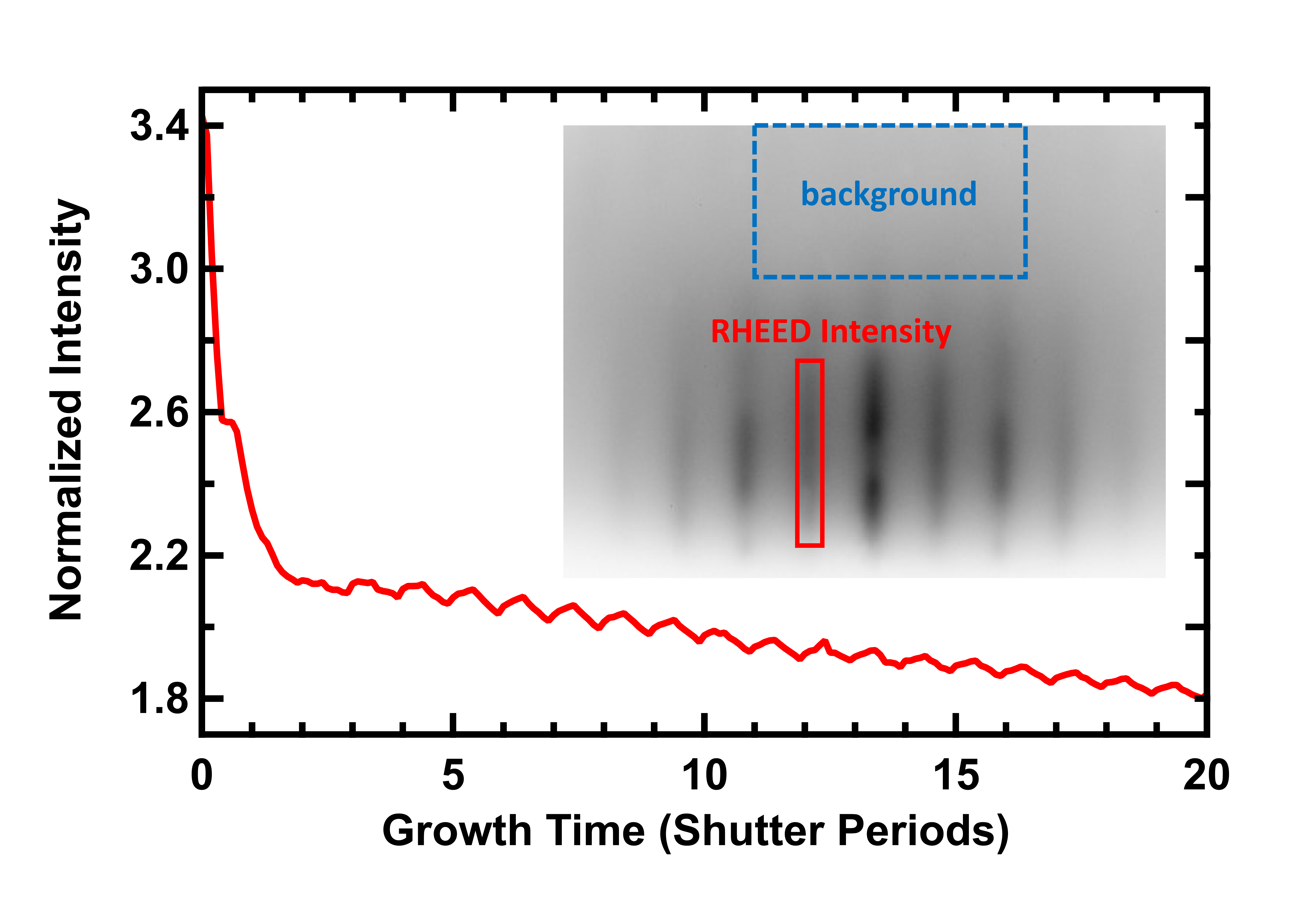}
    }
    \caption{(a) RHEED patterns for the Al$_2$O$_3$(0001) substrate, 5 nm Pt film, and 20 nm Fe$_3$Sn$_2$ film measured with the beam along $[1\bar{1}00]$ (left column) and $[11\bar{2}0]$ (right column) directions of the substrate. (b) Oscillations in the normalized RHEED intensity as a function of time. The RHEED intensity is measured within the red box and normalized by the background in the blue box.} 
\end{figure}

Fe$_3$Sn$_2$ is a ferromagnet with a high Curie temperature, $T_C$ = 670 K~\cite{giefers2006}, and saturation magnetization of 1.9 $\mu_B$ per Fe at low temperature~\cite{ye2018}.
Fig.~\ref{fig:kagome} shows the crystal structure of Fe$_3$Sn$_2$ (space group R$\bar{3}$m, with lattice constants \textit{a} = 5.338 \r{A} and \textit{c} = 19.789 \r{A}~\cite{giefers2006}) which consists of Fe$_3$Sn kagome layers and Sn spacer layers. 
In each Fe$_3$Sn monolayer, the Fe atoms form corner-sharing equilateral triangles surrounding hexagons, with Sn atoms sitting in the center of the hexagons.
The alternating sequence of one Sn$_2$ monolayer with honeycomb lattice and two Fe$_3$Sn kagome layers produces the layered crystal structure of Fe$_3$Sn$_2$.

Based on this layered structure, we synthesized Fe$_3$Sn$_2$ thin films on top of epitaxial Pt(111) buffer layers on Al$_2$O$_3$(0001) substates by atomic layer MBE.
The epitaxial growth was performed in an MBE chamber with a base pressure of $4\times10^{-10}$ Torr. 
Films were deposited on Al$_2$O$_3$(0001) substrates (MTI Corporation) prepared by annealing in air at 1000 $^\circ$C for 3 hours followed by annealing in ultrahigh vacuum (UHV) at 500 $^\circ$C for 30 minutes. 
A 5 nm Pt(111) buffer layer was deposited from an e-beam evaporator (Pt: 99.99\%, Kurt J. Lesker) onto the Al$_2$O$_3$ (0001) substrate by growing the first 0.6 nm at 440 $^\circ$C and the rest 4.4 nm while cooling down from 140 $^\circ$C to 80 $^\circ$C. 
The Pt buffer layer was post-annealed at 300 $^\circ$C to improve the crystallinity and surface roughness. 
The Fe$_3$Sn$_2$ layer was grown on Pt(111) at room temperature using the following atomic layer MBE sequence: deposit two atomic layers of Fe$_3$Sn with a Fe:Sn flux ratio of 3:1, deposit one atomic layer of Sn$_2$ with the growth time same as two Fe$_3$Sn layers, then repeat. 
The Fe and Sn fluxes were generated from Knudsen cells (Fe: 99.99\%, Alfa Aesar; Sn: 99.998\%, Alfa Aesar) and the growth rates were determined using a quartz deposition monitor that was calibrated by x-ray reflectometry. 
Typical growth rates were $\sim$ 0.85 \r{A}/min, $\sim$ 0.67 \r{A}/min, and $\sim$ 0.45 \r{A}/min for Fe, Sn, and Pt, respectively. 
To protect the sample from oxidation, a 3 nm Pt or 5 nm CaF$_2$ capping layer was deposited on top of the Fe$_3$Sn$_2$. 

RHEED patterns were measured during growth to
characterize the epitaxial growth and determine the in-plane lattice constants. 
Figure~\ref{fig:RHEED} shows the RHEED patterns for the Al$_2$O$_3$(0001) substrate (top row), 5 nm Pt buffer layer (middle row), and the Fe$_3$Sn$_2$ layer after 20 nm of growth (bottom row). 
The left and right columns show patterns taken for the beam along the $[1\bar{1}00]$ and $[11\bar{2}0]$ directions of the substrate, respectively.
With in-plane rotation of the sample, RHEED exhibits the same pattern every 60$^\circ$ (i.e. six-fold rotation symmetry) with the patterns alternating between $[1\bar{1}00]$-type and $[11\bar{2}0]$-type every 30$^\circ$. 
For the in-plane epitaxial alignment, the Pt(111) and Fe$_3$Sn$_2$(0001) surface unit cells are aligned with each other and rotated 30$^\circ$ with respect to the the Al$_2$O$_3$(0001) substrate, as $a_{Al_2O_3} \approx (\sqrt{3}/2)2a_{Pt} \approx (\sqrt{3}/2)a_{Fe_3Sn_2}$ (bulk values for in-plane lattice constant are $a_{Al_2O_3} = 4.759$ \r{A}, $2a_{Pt} = 5.549$ \r{A}, $a_{Fe_3Sn_2}=5.338$ \r{A}).

The streaky patterns observed during Fe$_3$Sn$_2$ growth signify diffraction from a two-dimensional surface. 
In addition, we observe oscillations (Fig.~\ref{fig:RHEED_oscillations}) in the normalized RHEED intensity where the maxima occurs for the Fe$_3$Sn termination and the minima occurs for the Sn$_2$ termination. 
The normalization is performed by dividing the intensity of the background and is helpful for canceling variations in the incident beam intensity and background lighting. 
Except for the change in RHEED intensity, we did not observe any other significant differences in the RHEED pattern between Sn$_2$ and Fe$_3$Sn terminations. Nevertheless, the presence of RHEED oscillations in atomic layer MBE confirms the modulation of the surface termination during growth.

\begin{figure}

\subfloat[\label{fig:XRD}]{
    \includegraphics[width=0.45\textwidth]{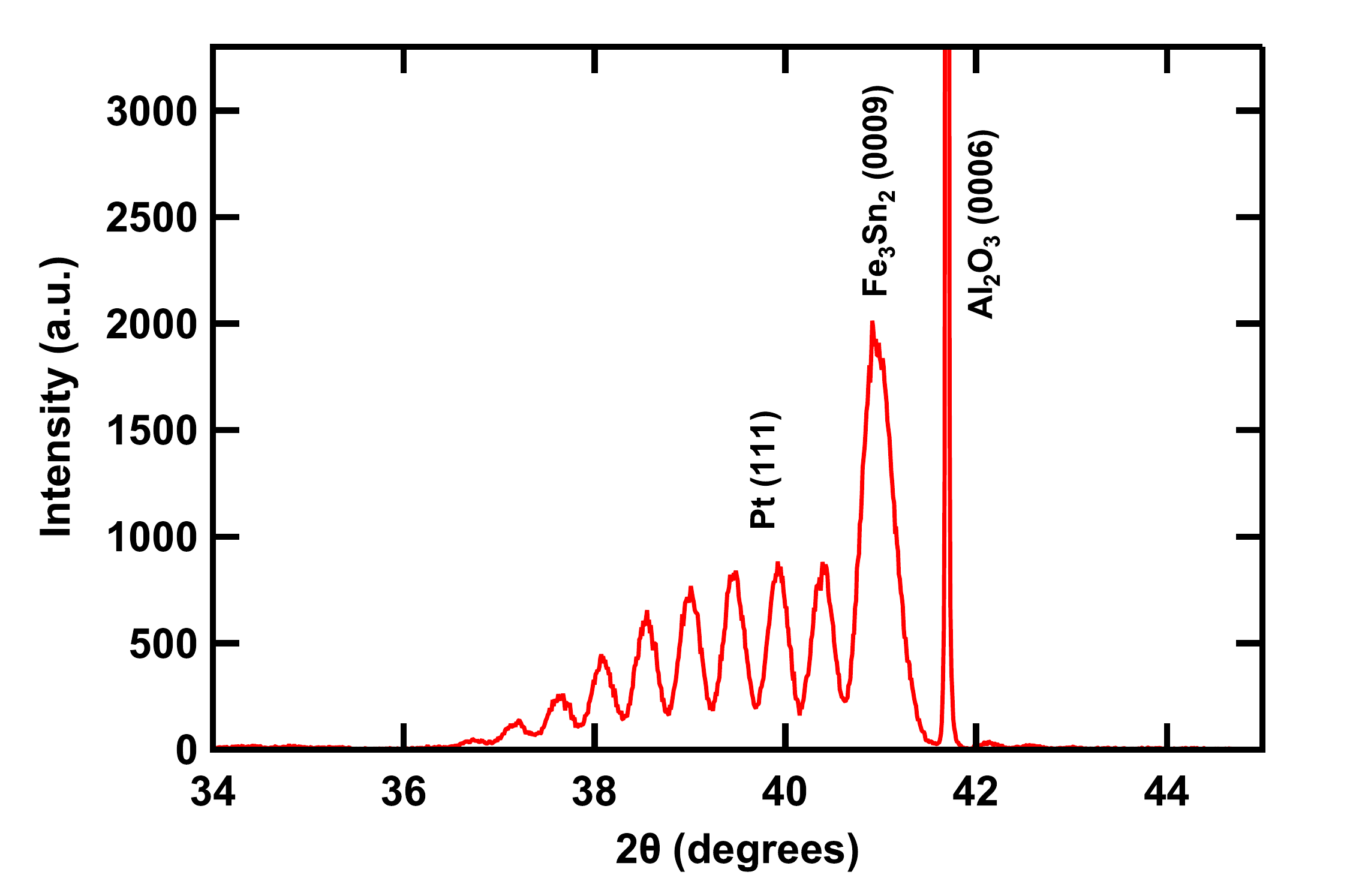}
}
  \hfill
\subfloat[\label{fig:rocking}]{
    \includegraphics[width=0.45\textwidth]{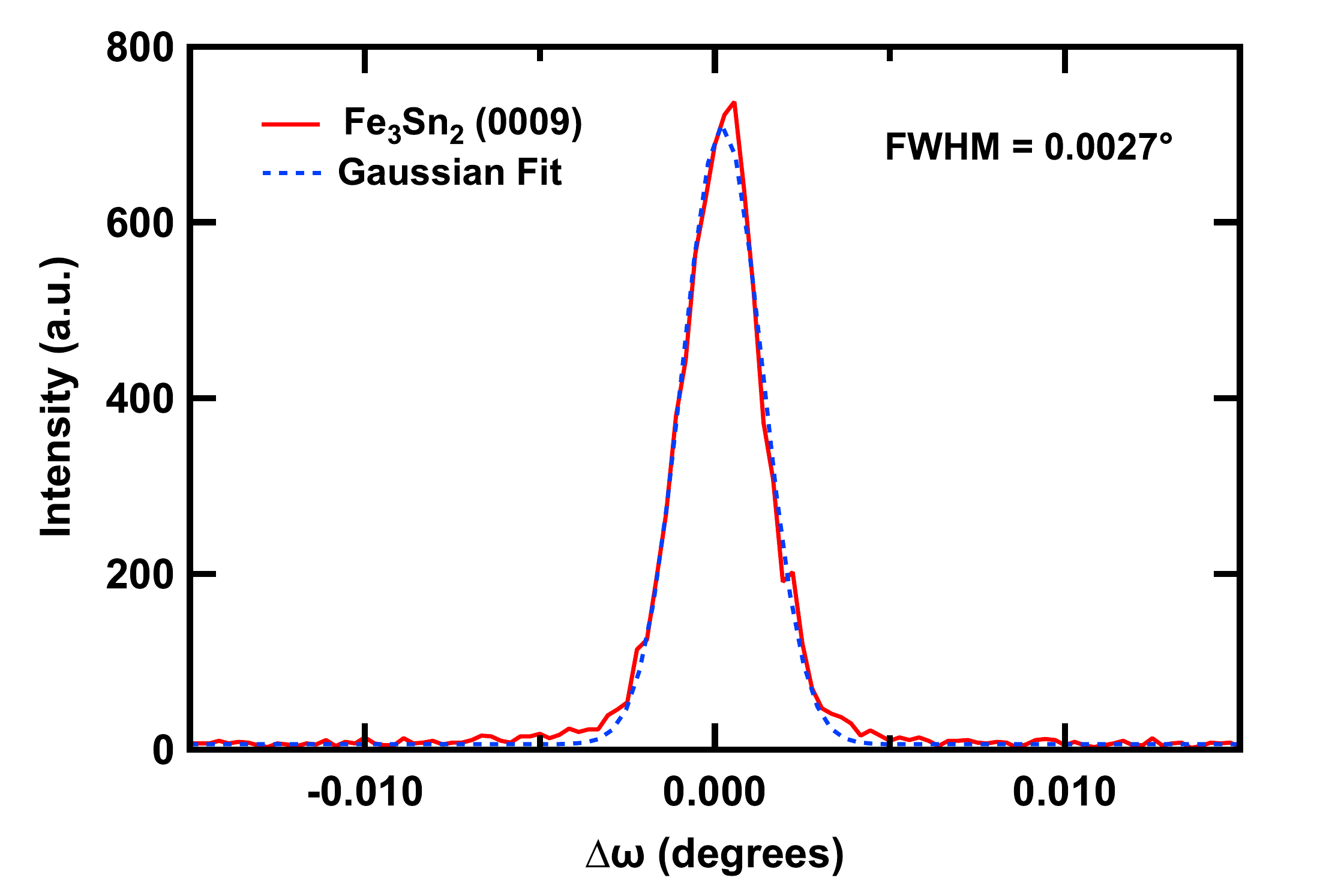}
} 
  \hfill
  \caption{(a) 2$\theta$-$\omega$ scan of a 20 nm Fe$_3$Sn$_2$ film grown on c-sapphire. (b) A triple axis $\omega$-relative scan (rocking curve) of the Fe$_3$Sn$_2$ (009) peak (red) with a Gaussian fitting (blue).}
\end{figure}

Films grown by this method were studied with XRD to analyze their crystal structure.
A representative $\omega$-$2\theta$ scan of a 20 nm film is shown in Figure~\ref{fig:XRD} and includes the Fe$_3$Sn$_2$ (0009) peak with several Laue-oscillations, indicating a high degree of film smoothness.
The out-of-plane lattice parameter extracted from analysis of this scan is 19.84 \r{A} which agrees well with previous reports of 19.789 \r{A} \cite{giefers2006}. A peak from the 5 nm Pt(111) buffer layer produces a shoulder on the Fe$_3$Sn$_2$ peak.
Larger range scans do not show additional peaks from the Fe$_3$Sn$_2$ or any other materials.
To further analyze the crystallinity of our films, rocking curve scans about the Fe$_3$Sn$_2$ (0009) peak were taken and analyzed.
By scanning the sample angle $\omega$ while keeping the detector fixed, this characterizes the angular distribution of the (0001) orientation relative to the film normal within the x-ray spot.
Figure~\ref{fig:rocking} contains a rocking curve or $\omega$-relative scan of the Fe$_3$Sn$_2$ using a triple-axis analyzer to achieve high resolution of the desired peak.
Fitting the peak with a standard Gaussian yields a full width half maximum (FWHM) of 0.0027$^\circ$.
Such a sharp peak indicates that films grown by atomic layer MBE have excellent mosaicity.

\begin{figure}
    \subfloat[\label{fig:MOKE}]{
   \includegraphics[width=0.46\textwidth]{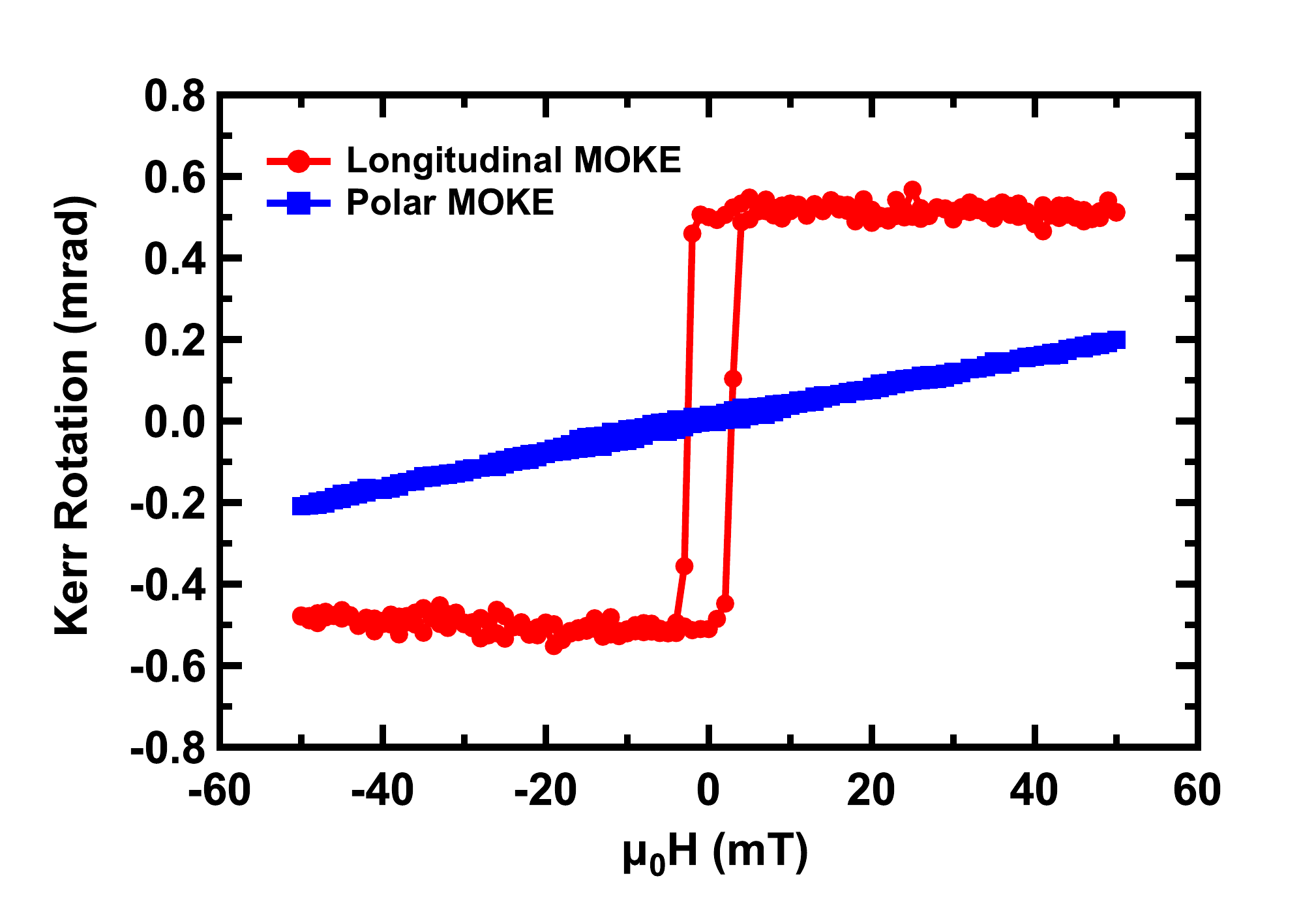}
}
  \hfill

   \subfloat[\label{fig:MOKEseries}]{
    \includegraphics[width=0.43\textwidth]{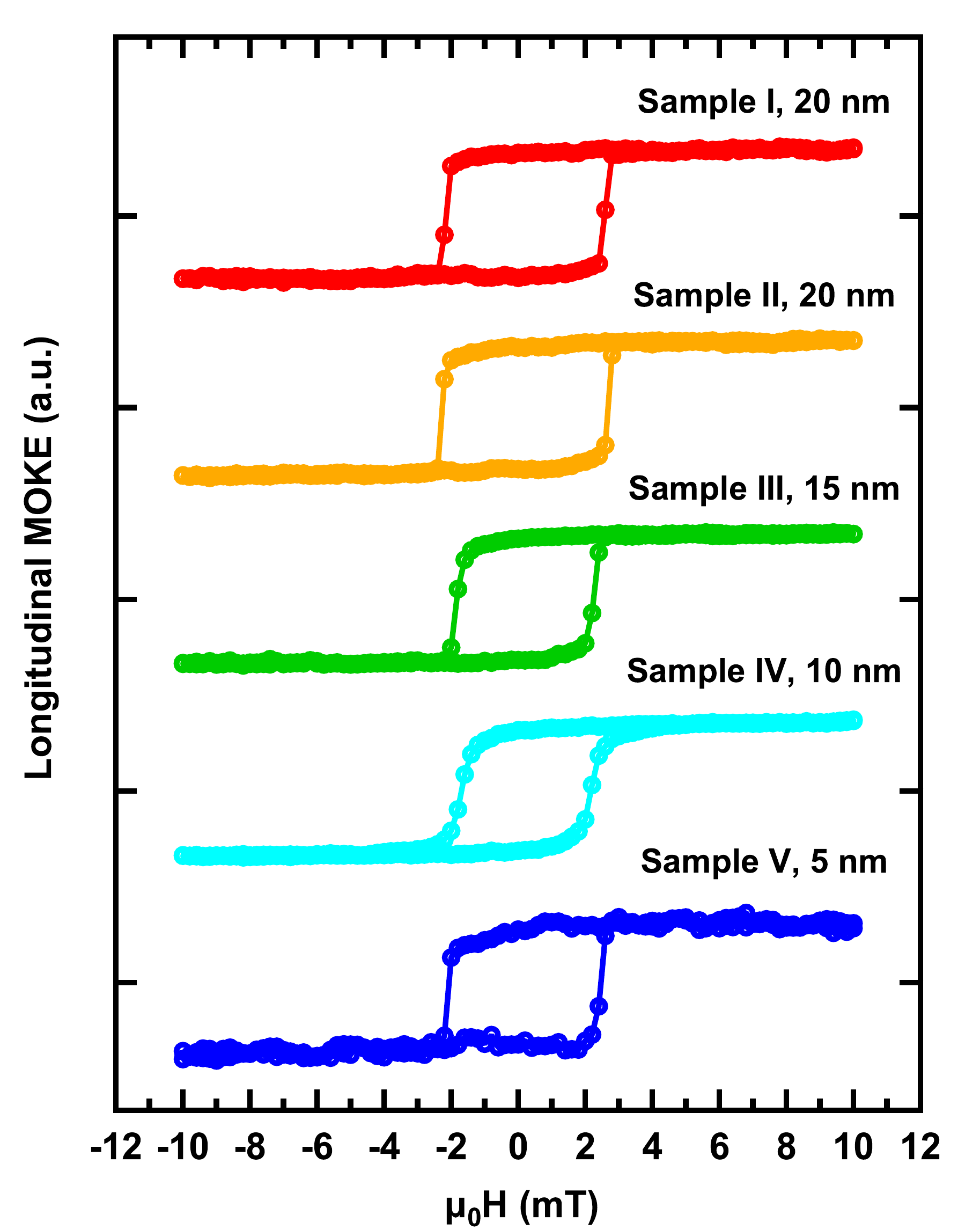}
    }
  \caption{(a). Longitudinal (red) and polar (blue) MOKE hysteresis loops of a 20 nm Fe$_3$Sn$_2$ film (Sample I). (b). Zoomed-in scans for five different samples with thicknesses ranging from 5 nm to 20 nm, showing similar magnetic properties. The data are normalized to the saturation values and offset for clarity.}
    \end{figure}

\begin{figure*}
    \subfloat[\label{fig:ANE_setup}] {
    \includegraphics[width=0.40\textwidth]{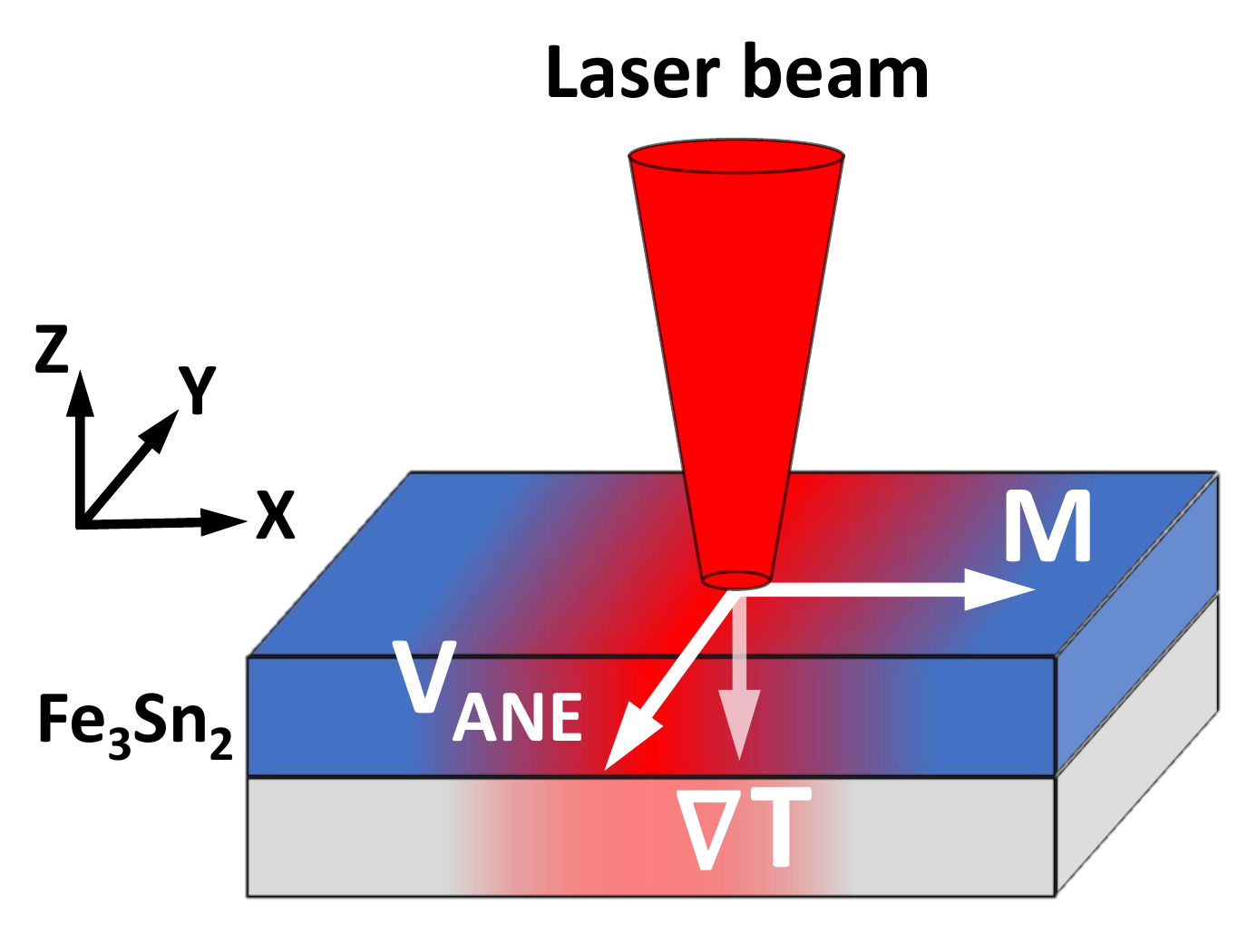}
    }
  \hfill
    \subfloat[\label{fig:ANE_device}] {
    \includegraphics[width=0.14\textwidth]{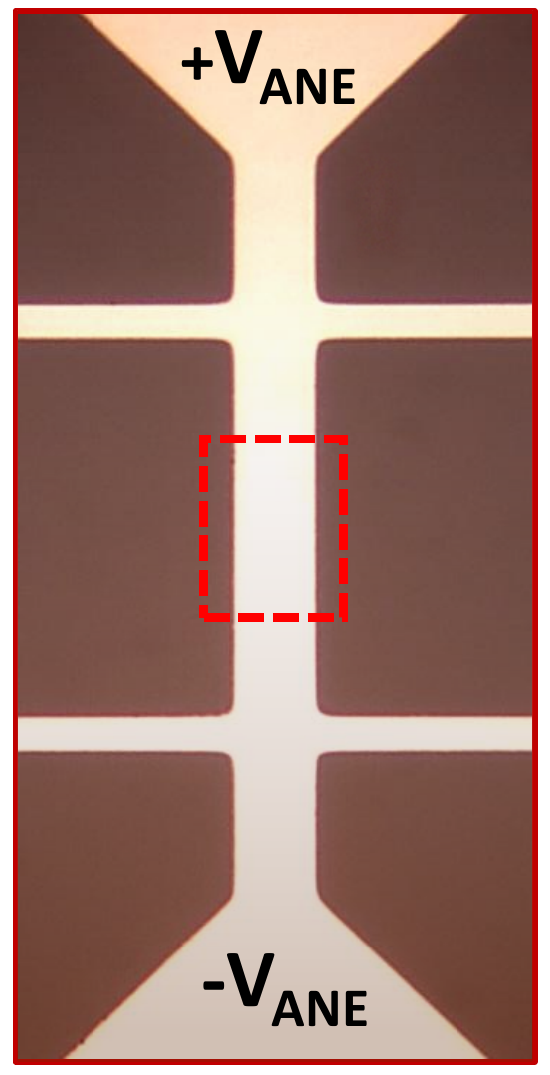}
    }
  \hfill
    \subfloat[\label{fig:ANE_hloop}]{
    \includegraphics[width=0.35\textwidth]{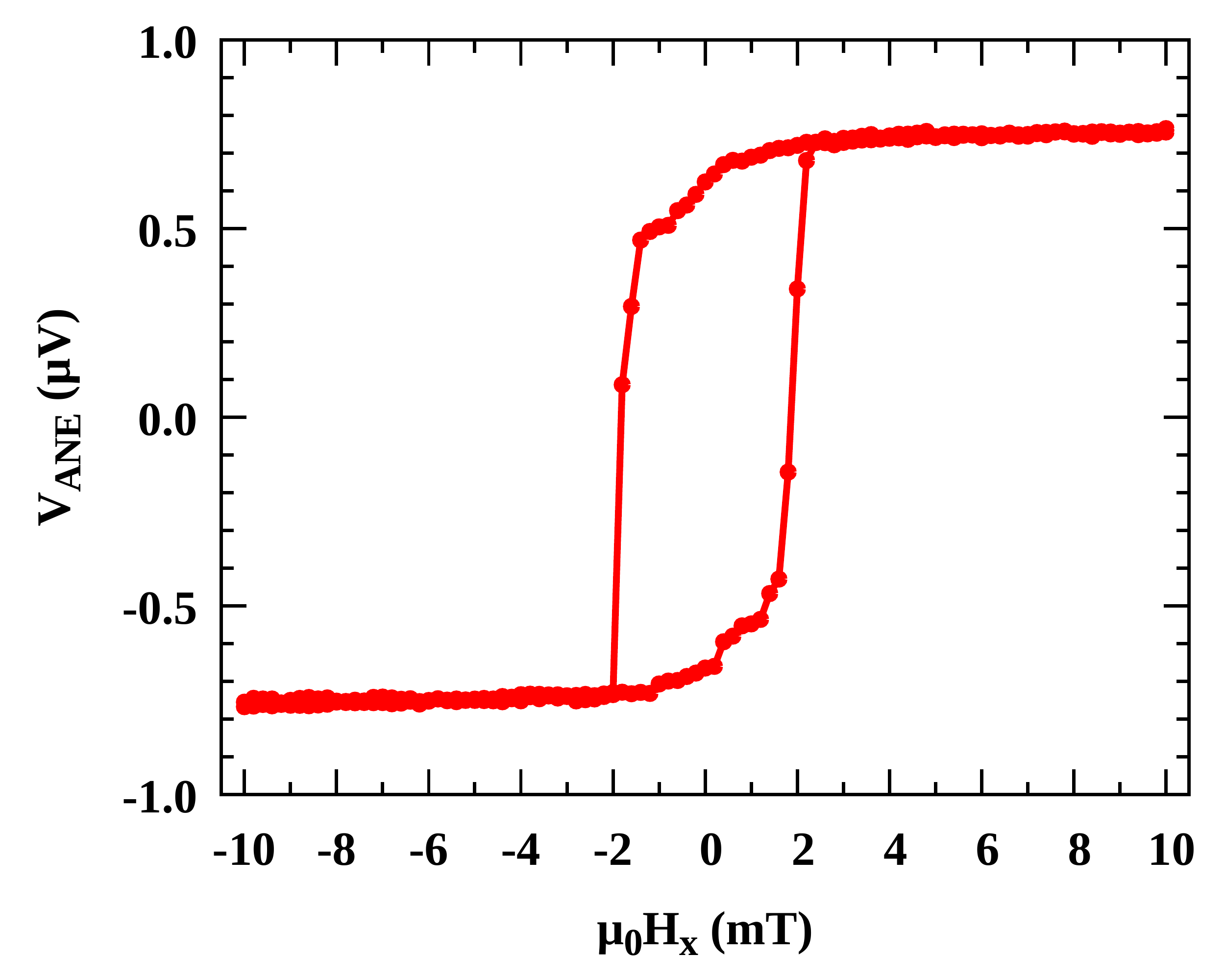}
    }
  \hfill
\subfloat[\label{fig:ANE_imaging}] {
    \includegraphics[width=\textwidth]{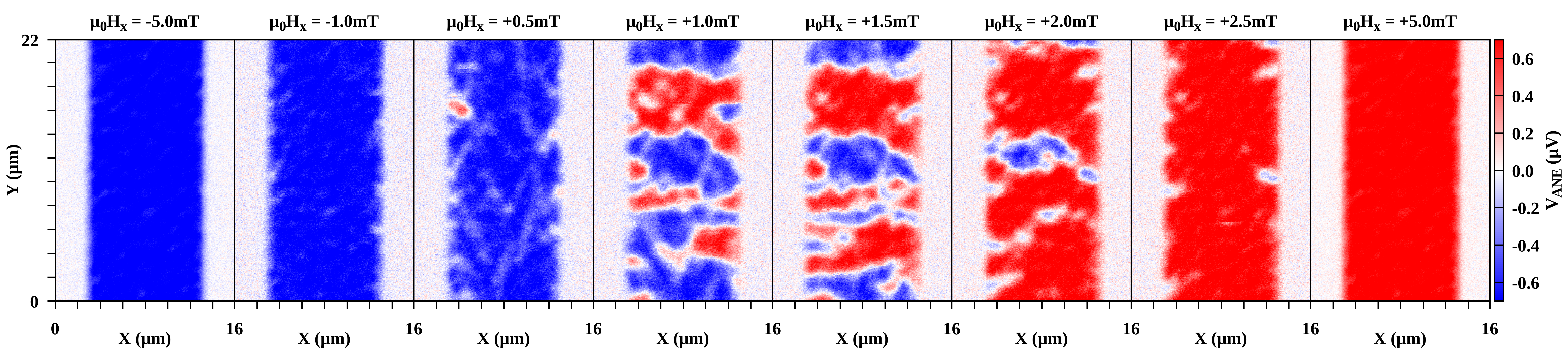}
    }
  \caption{
  (a) Schematics of thermal gradient microscopy.
  The laser beam is scanned over the sample surface, the induced local ANE voltage reflects the local magnetic properties.
  (b) Microscope image of a typical device, (the dashed rectangle corresponds to the area imaged in (d))
  (c) ANE hysteresis loop of a 20 nm Fe$_3$Sn$_2$ film.
  (d) Magnetization reversal of a 20 nm film through multidomain state imaged by ANE at a series of magnetic fields $\mu_0$H$_x$.}
\end{figure*}

To investigate the in-plane and out-of-plane magnetic properties of the Fe$_3$Sn$_2$ films, we measured longitudinal and polar MOKE hysteresis loops. 
The samples were probed using a linearly-polarized He-Ne laser (633 nm wavelength, $\sim100~\mu$W power, $\sim100~\mu$m spot size) and a polarizing beamsplitter, photodiode bridge, and lock-in amplifier (463 Hz intensity modulation) to detect the Kerr rotation. The laser beam had a $\sim$45$^\circ$ angle of incidence for longitudinal MOKE and normal incidence for polar MOKE.
Figure~\ref{fig:MOKE} shows a representative longitudinal hysteresis loop (red curve) measured on a 20 nm thick Fe$_3$Sn$_2$ sample (Sample I). 
The square hysteresis loop with a coercivity of 2.4 mT indicates ferromagnetic order with in-plane magnetization. 
In contrast, the polar hysteresis loop (blue curve) shows a small Kerr rotation with slight variation with out-of-plane magnetic field. Together, the longitudinal and polar MOKE loops show that the Fe$_3$Sn$_2$ samples have an easy-plane magnetic anisotropy.
This agrees with a previous study of Fe$_3$Sn$_2$ films grown by sputter deposition~\cite{khadka2020}.
To check the consistency of the synthesis and magnetic properties, we investigated additional samples grown by the same method with thickness varying from 5 to 20 nm. 
The results are shown in Fig.~\ref{fig:MOKEseries}.
All the samples show very similar magnetic properties, with square in-plane hysteresis loops and similar coercive fields.

The magnetic domain structure of Fe$_3$Sn$_2$ films are of interest due to the observation of skyrmions in bulk Fe$_3$Sn$_2$, but has not yet been studied in thin films. 
Longitudinal MOKE microscopy with oblique
angle incidence can detect the in-plane magnetization and therefore determine in-plane domain structure of our Fe$_3$Sn$_2$ films. However, in this manuscript, we choose to 
use thermal gradient microscopy (TGM)~\cite{weiler2012,gray2019,reichlova2019} over longitudinal MOKE to image domain structure because we found that it has a better signal-to-noise ratio in our experimental setup.

TGM is based on moving a laser spot over the sample surface, and recording a voltage induced by the local laser heating.
The thermal gradient generated in the out-of-plane direction $Z$ and a component of magnetization in the $X$ direction give rise to the anomalous Nernst effect, which is detected as a voltage along the $Y$ direction, $V_{ANE} \sim \left[ \nabla T \times \mathbf{M} \right]$ (see Fig.~\ref{fig:ANE_setup}).

For the ANE imaging, we fabricated 10 $\mu$m wide Hall bar devices by a combination of photolithography and argon ion milling (Fig.~\ref{fig:ANE_device}). The laser excitation for the thermal gradient was produced by a frequency-doubled
(BaB$_2$O$_4$ crystal) 
mode-locked Ti:Sapphire laser for a wavelength of 400 nm. The laser beam with 0.7 mW power was focused by a 50$\times$ objective lens (NA of 0.6) to a spot size of 0.9 $\mu$m, and a fast steering mirror in the 4f alignment scheme was used for scanning the laser spot over the sample surface. The intensity of the beam was modulated at a frequency of 120 kHz and the generated ANE voltage was detected using a lock-in amplifier.

We first utilized the ANE microscope to measure a detailed hysteresis loop at a fixed position. As shown in Fig.~\ref{fig:ANE_hloop} for magnetic field along the $X$ direction, the hysteresis loop shows a gradual reversal followed by a sharp switching behavior with coercivity of 1.9 mT. This has a similar coercivity but more gradual initial reversal than the in-plane hysteresis loops obtained by MOKE (Fig.~\ref{fig:MOKEseries}).

The origin of the different hysteresis properties is revealed by imaging the magnetic domain structure of Fe$_3$Sn$_2$ films at a series of magnetic fields. A representative sequence during the magnetization reversal is shown in Fig.~\ref{fig:ANE_imaging}.
Starting at -5.0 mT, the magnetization is in a saturated state along $-X$ (blue). 
The reversal initiates with the nucleation of white regions with $M_x \approx 0$, mainly at the edges of the sample. 
This can be explained by the minimization of domain wall energy as the edge boundary does not contribute a domain wall energy cost. 
The nucleation at the edges initiates magnetization reversal which results in a more rounded hysteresis loop compared to the uniform films.
With increasing magnetic field, domains of opposite polarity grow inward and coalesce across the channel. At about +1.0 mT, the magnetic structure is in a multidomain state with characteristic features (e.g. blue and red regions) ranging from 1 to 10 microns in size. By +2.0 mT, most of the magnetic moments have switched to $+X$ direction, with only a few regions remaining along $-X$. Finally, at +5.0 mT the magnetization reversal is complete and the films is fully saturated along $+X$.

In conclusion, we report the growth and characterization of kagome ferromagnet Fe$_3$Sn$_2$ thin films on Pt(111)/Al$_2$O$_3$(0001) by atomic layer molecular beam epitaxy.
Structural characterization by \textit{in situ} RHEED and XRD confirm the high quality of the epitaxial Fe$_3$Sn$_2$ films. The magnetic properties were investigated by magneto-optical Kerr effect and anomalous Nernst effect, confirming the easy-plane magnetic anisotropy of the thin films. 
Finally, the local magnetic structure was probed by ANE microscopy revealing the presence of in-plane oriented micrometer size domains during magnetization reversal.
These results highlight the potential for epitaxial growth to enable new scientific research in kagome magnets at the intersection of topology and magnetism.

\section*{Acknowledgements}

S.C., I.L., and R.K.K. acknowledge support from DARPA Grant No.~D18AP00008. A.J.B. and R.K.K. acknowledge support from DOE Grant No.~DE-SC0016379. This research was partially supported by the Center for Emergent Materials, an NSF MRSEC, under award number DMR-2011876. 
\\

\section*{Author contributions}

S.C., I.L., and R.K.K. conceived the experiments. S.C. conducted the MBE growth and MOKE measurements. I.L. conducted the ANE measurements.
A.J.B. conducted the XRD measurements.
All authors participated in data analysis and preparation of the manuscript.

\bibliography{main.bib}

\end{document}